# Joint User Association and UAV Location Optimization for Two-Tired Visible Light Communication Networks

Alireza Qazavi
*Department of Electrical and Computer Engineering*
*Isfahan University of Technology*
Isfahan, Iran
a.qazavi@ec.iut.ac.ir

Foroogh S. Tabataba
*Department of Electrical and Computer Engineering*
*Isfahan University of Technology*
Isfahan, Iran
fstabataba@iut.ac.ir

Mehdi Naderi Soorki
*Engineering Faculty*
*Shahid Chamran University of Ahvaz*
Ahvaz, Iran
m.naderisoorki@scu.ac.ir

*Abstract*— **In this paper, an unmanned aerial vehicle (UAVs)-assisted visible light communication (VLC) has been considered which has two tiers: UAV-to-centroid and device-to-device (D2D). In the UAV-to-centroid tier, each UAV can simultaneously provide communications and illumination for the centroids of the ground users over VLC links. In the D2D tier, the centroids retransmit received data from UAV over D2D links to the cluster members. For network, the optimization problem of joint user association and deployment location of UAVs is formulated to maximize the received data, satisfy illumination constraint, and the user cluster size. An iterative algorithm is first proposed to transform the optimization problem into a series of two interdependent sub problems. Following the smallest enclosing disk theorem, a random incremental construction method is designed to find the optimal UAV locations. Then, inspired by unsupervised learning method, a clustering algorithm to find a suboptimal user association is proposed. Our simulation results show that the proposed scheme on average guarantees the users brightness 0.77 lux more than their threshold requirements. Moreover, the received bitrate plus number of D2D connected users under our proposed method is 50.69% more than the scenario in which we have RF Link instead of VLC link and do not optimize UAV location.**

## I. Introduction

Recently, a wide range of UAV applications such as security, control, surveillance, and reconnaissance of ground areas, in upcoming telecommunication networks has been proposed which are difficult to reach quickly without drones [1]. In addition to assisting smart cities in meeting public security needs, UAVs can also provide support for the delivery of medical supplies and the transport of blood and pharmaceutical products, as well as emergency management such as forest fires, protection, and inspection of critical infrastructure, coastal surveillance and police reinforcement. UAVs are also used for military and electronic warfare applications, or for traffic control applications. Reference [1] provides an overview of the various applications of drones in the smart city and public safety.

Providing an in-flight high-speed internet connection with such Aerial–IoT vehicles is becoming more challenging than it is today. This is because the wireless radio spectrum is extremely congested due to the high density of the devices and their high power requirements. In this regard, VLC, is an attractive offer for connecting UAVs to users, because of its highly directional communications, VLC can activate its multicast links from a single UAV, which can be very challenging if using wireless radio communication due to signal interference [2].

There are several recent works such as [3]–[6], and [7] that consider the simultaneous use of VLC and UAV communication technologies. In [7], the authors presents the first use of UAVs and VLC technology with the aim of optimizing power consumption. [3] Suggests optimizing the position of UAVs providing a reliable communication links between UAVs and vehicles. The works in [4] and [5] also focused on collaborating non-orthogonal multiple access technologies and machine learning with VLCs. In [7], the authors examines a wide range of UAV and VLC applications such as lighting of passages at night or guiding the elderly. They minimize power consumption and does not concern about data rate. In addition, they optimize UAV location and do not optimize UAV association and D2D connections. However, in [8], the authors consider D2D and user association problem but do not consider UAV location optimization. However, none of the existing works in [3]–[6] and [7] does not consider joint user clustering and optimal location problem in the UAV-assisted VLC network, especially when D2D communication is possible between users of each cluster.

The main contribution of this paper is to jointly optimize the user association and UAV location in the UAV-assisted VLC networks. In particular, we consider a two-tiered network in which a number of UAVs must provide data transmission and illumination to the centroid of users over VLC links, and then the user centroid retransmit the data to their members over short range D2D links. We transform this NP-hard problem to two interdependent sub problems. Based on the smallest enclosing disk theorem [9], we design a random incremental construction method to find the optimal UAV locations and meet the minimum brightness requirement of centroids. Particularly, we show that for each fixed user clustering, there is a unique optimal position for UAV, which is in the center of the smallest disk that covers all users in the UAV lighting area. Then, using unsupervised learning method, we propose a clustering algorithm to find a suboptimal user association. Our simulation results show that the proposed scheme on average guarantees the users brightness 0.77 lux more their threshold requirements. Moreover, the received bitrate under our proposed method is 50.69% more than the scenario in which we have RF Link instead of VLC link and do not optimize UAV location that proposed in [8].

The rest of the paper is organized as follows. The system model and problem formulation is presented in Section II. In Section III, our proposed two-step method for user association and UAV location optimization is discussed. The simulation results are provided in Section IV. Finally, conclusions are drawn in Section V.

## II. SYSTEM MODEL AND PROBLEM FORMULATION

### A. UAV-cluster communication

We cconsider a given the geographical area in which the two UAVs, each capable of serving up to a maximum of K users, provide downlink communications. As shown in Figure 1, UAVs provide both instantaneous illumination of the area and transmission of information to a target area. This VLC network consists of N ground users and two UAVs that act as base stations (BSs).

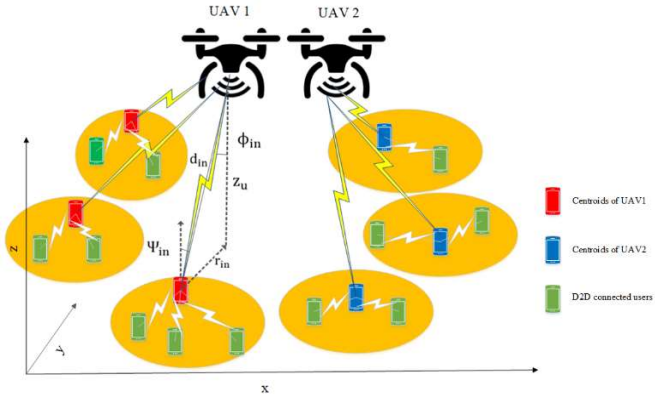

*Figure 1:System model scenario*

Let $\mathcal{N} = \{1,2, \dots N\}$ be a set of ground users that are uniformly distributed on a ground. Some users are directly connected to the UAVs. While the rest of the users are indirectly connected to the UAVs using D2D communication with drone-connected users. Considering the directional transmission feature of VLCs, we assume that each user is connected to a maximum of one UAV. The channel gain of a VLC link can be expressed as follows [7]:

$$h_{in} = \begin{cases} \frac{(m+1)A}{2\pi d_{in}^2} g(\Psi_{in}) \cos^m(\phi_{in}) \cos(\Psi_{in}), & 0 < \Psi_{in} < \Psi_c, \\ 0, & \Psi_{in} > \Psi_c, \end{cases} \quad (1)$$

Where $h_{in}$ represents the channel gain between the UAV i and the user n, $A$ represents the detector area, and $d_{in}$ represents the distance between the UAV I and user n. In addition, $m = -\ln 2 / \ln(\cos \Phi_{1/2})$, where $\Phi_{1/2}$ is the transmitter half angle, $\Psi_{in}$ is the collision angle, $\phi_{in}$ is the radiation angle, $\Psi_c$ is half of the field of view (FoV) in the receiver, and $g(\Psi_{in})$ is the optical focusing gain. Given by the following equation:

$$g(\Psi_{in}) = \begin{cases} \frac{n_r^2}{\sin^2 \Psi_c}, & 0 < \Psi_{in} < \Psi_c, \\ 0, & \Psi_{in} > \Psi_c, \end{cases} \quad (2)$$

Where $n_r$ is the refractive index.

For simplicity, we assume that the altitude of both UAVs is equal to $z_u = h$ and constant. The initial positions of the UAVs $(\boldsymbol{w}_{ui}^0(1), \boldsymbol{w}_{ui}^0(2), z_u)$, i = 1,2 and the user positions $(\boldsymbol{w}_n(1), \boldsymbol{w}_n(2), 0)$, $n \epsilon \mathcal{N}$ (where $\boldsymbol{w}_n$ and $\boldsymbol{w}_{ui}$ ($i = 1,2$) are given by the horizontal position vectors of the user n and the UAV i, respectively. Thus, we have $d_{in} = \sqrt{(\boldsymbol{w}_{ui}^0(1) - \boldsymbol{w}_n(1))^2 + (\boldsymbol{w}_{ui}^0(2) - \boldsymbol{w}_n(2))^2 + h^2}$ and $\cos \phi_{in} = \cos(\Psi_{in}) = \frac{z_u}{d_{in}} = \frac{h}{d_{in}}$.

Note that these angles are based on the assumption that the transmitters and receivers are vertically downward and upward, respectively. Here, cameras or photodiodes can be placed as receivers on ground users to convert light signals into electrical signals and support communications for mobile users [7].

We define the binary variables $\{x_{1n}\}$ and $\{x_{2n}\}$. As representing the allocation of the UAV to the user. If $x_{1n} = 1$, the first UAV serves the user n, otherwise $x_{1n} = 0$. Similarly, if $x_{2n} = 1$, the second UAV serves the user n, otherwise $x_{2n} = 0$. Assume that each UAV supports a maximum of K users, and that each user can only be served by a maximum of one UAV, which results in the following limitations: $\sum_{n \epsilon \mathcal{N}} x_{1n} \leq K$, $\sum_{n \epsilon \mathcal{N}} x_{2n} \leq K$, $x_{1n} + x_{2n} \leq 1$. $\forall n \epsilon \mathcal{N}$.

If the user n is served by the UAV i, that is, if $x_{in} = 1 (i = 1,2, n \epsilon \mathcal{N})$, a lower bound for the channel capacity (C) of the VLC link between the UAV i and the user n is given below [10].

$$C \geq \frac{1}{2} \log_2 \left(1 + \frac{e}{2\pi} \left(\frac{\zeta p_u h_{in}}{\sigma_w}\right)^2\right). \quad (3)$$

Where $e$ is the Euler parameter, $\sigma_w$ is the Gaussian noise standard deviation, $\zeta$ is the luminance target coefficient which is numerically between 0 and 1, and $p_u$ is the nominal LED light power of each UAV.

Due to the use of channels, the interference between users connected to the same UAV is not considered. In addition, there is no interference between UAVs because in VLC, the signal transmission is directional, and there is enough large distance between ground coverage areas of UAVs. Note that LED UAVs are more likely to be deployed in dark scenarios (such as at night) [6]. Since VLC provides both communication and brightness, brightness limits must also be considered. The luminous flux density indicates the luminosity of the luminous surface, which corresponds to $\eta_n = \zeta p_u h_{in}$ [11]. The position of the UAV can be precisely adjusted to reduce the distance between the UAVs and the users, and thus increase the transmission rate.

Accordingly, the unit achievable rate (bps / Hz) for the user n associated with the two UAVs can be calculated as follows.

$$R_n = x_{1n}\frac{1}{2}\log_2\left(1 + \frac{e}{2\pi}\left(\frac{\zeta p_u h_{1n}}{\sigma_w}\right)^2\right) \\ + x_{2n}\frac{1}{2}\log_2\left(1 + \frac{e}{2\pi}\left(\frac{\zeta p_u h_{2n}}{\sigma_w}\right)^2\right). \forall n \epsilon \mathcal{N}. \quad (4)$$

Therefore, the total rate of users who are served by two UAVs can be expressed as

$$R_{sum}^u = \sum_{n\epsilon\mathcal{N}} R_n. \quad (5)$$

*B. D2D tier*

Over D2D tier, the directly connected users to the UAVs (centroids) can provide service for the other users locating in the D2D communication range. We define binary variables $\{y_{nk}\}$. If the user n can get the content from the user k (centroid) via a D2D connection, $y_{nk} = 1$; otherwise $y_{nk} = 0$. Only when the k user's service is served by one of the UAVs, he/she can have a chance to provide wireless service to the user n ($n \neq k$) via D2D connection. That is, if $x_{1n} + x_{2n} = 0$, then $y_{nk} = 0, \forall k\epsilon\mathcal{N}, n \neq k$. If $x_{1n} + x_{2n} = 1$, then the value of $y_{nk}$ may be 0 or 1. In summary, we have the following limitations.

$$y_{nk} \leq x_{1k} + x_{2k}, \forall n, k\epsilon\mathcal{N}, n \neq k. \quad (6)$$

It is assumed that, at any given time, each user is only served by a maximum of one transmitter (a specific UAV or one of the users serviced by the UAVs), which results in the following limitations.

$$x_{1n} + x_{2n} + \sum_{k\epsilon\mathcal{N}} y_{nk} \leq 1, \forall n\epsilon\mathcal{N}. \quad (7)$$

When the user n is served by the user k with a D2D connection, $y_{nk} = 1$.

We assume that users directly served by UAVs only connect to users via D2D connections if the distance between them is less than or equal to $d_{D2D}$ which is reliable communication range. Therefore, we obtain the constraint as $y_{nk} = 0$ if $d_{nk} \geq d_{D2D}$.

*C. Problem formulation*

we define $N_k, k\epsilon\mathcal{K}$ as the number of users that can be serviced by user k. We have: $N_k = \sum_{n=1,n\neq k}^{n=N} y_{nk}$.

In order to improve the performance of UAV-served users and to ensure that more users can have a chance of being served via D2D connections, we add the total weighted rate of UAV-serviced users and the total number of served users. Therefore, the optimization problem can be formulated as follows:

$$\max_{w_{u1},w_{u2},x_1,x_2,Y} a \sum_{n\epsilon\mathcal{N}}((x_{1n}\frac{1}{2}\log_2\left(1 + \frac{e}{2\pi}\left(\frac{\zeta p_u h_{1n}}{\sigma_w}\right)^2\right) \\ + x_{2n}\frac{1}{2}\log_2\left(1 \\ + \frac{e}{2\pi}\left(\frac{\zeta p_u h_{2n}}{\sigma_w}\right)^2\right)) + b\sum_{j\epsilon\mathcal{K}} N_j \quad (8.a)$$

$$s.t. \; x_{1n} + x_{2n} + \sum_{k\epsilon\mathcal{N}} y_{nk} \leq 1, \forall n\epsilon\mathcal{N}, \quad (8.b)$$

$$\sum_{n\epsilon\mathcal{N}} x_{1n} \leq K, \sum_{n\epsilon\mathcal{N}} x_{2n} \leq K, \quad (8.c)$$

$$y_{nk} \leq x_{1k} + x_{2k}, \forall n, k\epsilon\mathcal{N}, n \neq k. \quad (8.d)$$

$$y_{nk} = 0, \; if\, d_{nk} \geq d_{D2D}, \forall n, k\epsilon\mathcal{N}, \quad (8.e)$$

$$\zeta p_u h_{in} \geq \eta_{th}, \quad (8.f)$$

$$x_{1n}\epsilon\{0,1\}, x_{2n}\epsilon\{0,1\}, \quad \forall n\epsilon\mathcal{N}, \quad (8.g)$$

$$y_{nk}\epsilon\{0,1\}, y_{nn} = 0, \quad \forall n\epsilon\mathcal{N}. \quad (8.h)$$

Where $\boldsymbol{x_1} = [x_{11}, x_{12}, ..., x_{1N}]^T$, $\boldsymbol{x_2} = [x_{21}, x_{22}, ..., x_{2N}]^T$, and $\boldsymbol{Y} = [y_{n,k}]_{N\times N}$. $\eta_{th}$ In (8.f) is the brightness threshold of the receiver. $a$ in (8.a) is the weight assigned to the total user rates serviced by UAVs. $b$ in (8.a) is the weight assigned to the total number of users served via D2D links. Where $0 \leq a \leq 1, 0 \leq b \leq 1$, and $a + b = 1$. When $a \gg b$, the problem in (8) is simply simplified by maximizing the total user rate - served by UAVs. Conversely, if $a \ll b$, the problem in (8) only changes to maximize the total number of users served via D2D connections.

The problem in (8) is NP-hard, because it is a mixed integer optimization problem that includes binary variables $\boldsymbol{x_1}, \boldsymbol{x_2}$ and $\boldsymbol{Y}$ and non-integer variables [12]. We can get the optimal global value through the comprehensive search method. However, its computational complexity is extremely high, even in moderate-scale networks. By the way, the problem (8) is challenging due to the interdependence between $(\boldsymbol{w_{ui}}(1), \boldsymbol{w_{ui}}(2), z_u), i = 1,2$, $\boldsymbol{Y}, \boldsymbol{x_1}, \boldsymbol{x_2}$. Therefore, further manipulation is needed to find the answer. Note that by specifying the binary variables $\{x_{1n}\}$ and $\{x_{2n}\}$, Y is uniquely specified. Therefore, Y is an auxiliary variable.

III. THE PROPOSED SUB-OPTIMAL JOINT USER ASSOCIATION AND UAV LOCATION

First, we assume that the cell allocation is fixed, indicating that the centroids are known. According to this assumption, the binary variables $\{x_{1n}\}$ and $\{x_{2n}\}$ Will be fixed and known, and therefore, Y will be fixed and known accordingly. Therefore, $\sum_{j\epsilon\mathcal{K}} N_j$ will also be constant. In addition, under this assumption, the optimization problem for each UAV becomes independent of the other UAV. In addition, a UAV only needs to consider the farthest user connected to it. As soon as the needs of the farthest user are met, the needs of all other users will be met. Based on this insight, the optimization problem for each UAV i can be expressed as follows:

$$\max_{w_{ui}} \log_2\left(1 + \frac{e}{2\pi}\left(\frac{\zeta p_u h_{in_i^*}}{\sigma_w}\right)^2\right) \qquad (9.a)$$

$$s.t. \quad \zeta p_u h_{in_i} \geq \eta_{th} \quad \forall n_i \epsilon \{n \epsilon \mathcal{N} | x_{in} = 1\}, \qquad (9.b)$$

Where $n_i$ represents the users are connected to the UAV $i$, and $n_i^*$ represents the farthest user from the UAV $i$. Note that the farthest user changes dynamically as the UAV changes position. The optimal position of each UAV can be determined based on the following lemma.

**Lemma1.** For a fixed cell assignment, there is a unique optimal position for UAV i in the center of the smallest disk that covers all users in the UAV i lighting area.

**Proof.** By substituting (1) in (9.b), we obtain:

$$d_{in_i} \leq \sqrt[m+3]{\frac{p_u}{V}} \quad \forall n_i \epsilon \{n \epsilon \mathcal{N} | x_{in} = 1\}, \qquad (10)$$

In which $V = \frac{2\pi \eta_{th}}{(m+1)Ag(\Psi_{in_i^*})z_u^{m+1}\xi}$. Thus, (10) provides a high boundary for $d_{in_i}$. According to the previous description, it is sufficient that the farthest user associated with UAV $i$ is at a distance less than $\sqrt[m+3]{\frac{p_u}{V}}$ from that UAV, in which case all users that is connected to UAVs meet condition (10). according to (1) and the relation of the inverse $h_{in_i^*}$ to the square $d_{in_i^*}$, as well as due to the ascending $\log_2\left(1 + \frac{e}{2\pi}\left(\frac{\zeta p_u h_{in_i^*}}{\sigma_w}\right)^2\right)$ by increasing $h_{in_i^*}$, (with constant $p_u$, $\zeta$ and $\sigma_w$) we have: $\arg\max_{w_{ui}} \log_2\left(1 + \frac{e}{2\pi}\left(\frac{\zeta p_u h_{in_i^*}}{\sigma_w}\right)^2\right) = \arg\min_{w_{ui}}\left(d_{in_i^*}\right)$. Which $\arg\min_{w_{ui}}\left(d_{in_i^*}\right) = \arg\min_{w_{ui}}\left(r_{in_i^*}\right)$. where $r_{in_i^*}$ is the horizontal distance between UAV $i$ and user $n_i^*$ on the two-dimensional x-y plane. $r_{in_i^*}$ is defined by $r_{in_i^*} = \sqrt{d_{in_i^*}^2 - z_u^2}$.

Condition (10) must also be met for all available users.

When the system parameters are given, the variable V will be constant. In addition, the variable $r_{in_i^*}$ will include the following constraint: $r_{in_i^*} \geq r_{in_i}$ for $n_i \epsilon \mathcal{U}_i$. where $n_i$ represents the user n to be served by UAV i and $\mathcal{U}_i$ is the set of users served by UAV i. Therefore, the problem can be rewritten as follows:

$$\arg\min_{w_{ui}}\left(r_{in_i^*}\right) \qquad (11.a)$$

$$d_{in_i} \leq \sqrt[m+3]{\frac{p_u}{V}}, \quad \forall n_i \epsilon \{n \epsilon \mathcal{N} | x_{in} = 1\}, \qquad (11.b)$$

$$r_{in_i^*} \geq r_{in_i}, \quad \text{for } n_i \epsilon \mathcal{U}_i. \qquad (11.c)$$

Without considering the condition (11.b) for the above problem, it can be said that the purpose of problem (11) is to find the optimal position of the disk center with radius $r_{in_i^*}$, related to all users that assigned to UAV $i$. we can show that a disk that contains all users with smaller disk exists [7]. This ends the proof. ∎

We use a random incremental algorithm to solve the problem (11) [9]. First, we generate a random permutation $w_1, \ldots, w_{N_i}$ from the position of the users in $\mathcal{U}_i$. Consider $\mathcal{U}_l \coloneqq \{w_1, \ldots, w_l\}$. While maintaining and considering the smallest $\mathcal{U}_l$ all-inclusive disk in each step $D_l$, we add the UAV position points one by one. Knowing the positions of the users, the smallest all-inclusive disk can be obtained by making a random increment if the brightness requirement of all users is met according to the condition (11.b). With the smallest disk, $D_{U_i}$, obtained from this algorithm, which condition (11.b) will be hold for the users inside this disk, the center of disc will be the optimal for the UAV i. Then, the total data receiving rate from UAV i will be obtained as follow:

$$R_{sum}^{u_i} = \sum_{n_i \epsilon \mathcal{U}_i}\left(\frac{1}{2}\log_2\left(1 + \frac{e}{2\pi}\left(\frac{\zeta p_u h_{in_i}}{\sigma_w}\right)^2\right)\right). \qquad (12)$$

If the disk is obtained and therefore the new position of the UAV fails to meet the requirement (11.b) for all connected users, we repeat the outermost loop until the limit (11.b) is met.

Since the distance between the UAV and the user depends not only on the location of the UAV but also on the location of the users, the next step is to allocate the cell to increase the total user reception rate along with the number of users using the D2D connection. To this end, in this sub problem, we consider the position of the UAVs ($w_{u1}$ and $w_{u2}$) constant and optimize the allocation of cells, $\{x_{1n}\}$, $\{x_{2n}\}$ and Y. In this case, we look at this problem as a clustering problem; hence, the problem (8) changes as follows.

$$\max_{x_1,x_2,Y} a \sum_{n \epsilon \mathcal{N}}\left(x_{1n}\frac{1}{2}\log_2\left(1 + \frac{e}{2\pi}\left(\frac{\zeta p_u h_{1n}}{\sigma_w}\right)^2\right)\right.$$
$$\left. + x_{2n}\frac{1}{2}\log_2\left(1 + \frac{e}{2\pi}\left(\frac{\zeta p_u h_{2n}}{\sigma_w}\right)^2\right)\right) \qquad (13.a)$$
$$+ b \sum_{j \epsilon \mathcal{K}} N_j$$

$$s.t. \quad (8.b), (8.c), (8.d), (8.e), (8.f), (8.g), (8.h)$$

It should be noted that the value of **Y** depends on the values of $x_1$ and $x_2$. As soon as $x_1$ and $x_2$ are fully defined. Y is uniquely determined by distance constraints. From this perspective, our goal is to right select the 2K of user so that the two UAVs serve those users in order to maximize the target value (13. a). This problem can be considered a clustering problem and can be solved with a cluster-based algorithm [8]. Users served by UAVs can be considered as cluster centers, and cluster users that apply within distance limits can be serviced by the corresponding cluster centers using a D2D connection. Therefore, we need to find the appropriate 2K cluster to maximize the total weighted rate of the cluster center users and the number of cluster users, which apply to distance constraints.

The problem of clustering is a problem of unsupervised learning, without a number of predetermined classes, in the field of machine learning. Data samples are divided into several different clusters based on some similarities. We know that the K-means clustering algorithm is a typical algorithm for solving unsupervised learning problems based on the iterative method, which is widely used due to its simplicity and less computational complexity. In the K-means clustering algorithm, first, the number K of the initial cluster center is randomly selected. Then, each sample is placed in a cluster that has the shortest Euclidean distance from the center of that cluster. Cluster centers, therefore, can be updated with the average sample values of each cluster, as long as the average values of the clusters do not change. Inspired by the K-means clustering algorithm, we present a similar cluster-based algorithm for the approximate answer to the problem (13).

First, we randomly select the number of 2K users as the primary cluster centers and specify their position with $\{C_1, C_2, ..., C_{2K}\}$.

Then, the distance between the user n and the center of the k cluster can be calculated for all ns and ks as follows.

$$d_{nk}^c = \|w_n - c_k\|, \forall n \in \mathcal{N}, \forall k = 1,2,...,2K. \quad (14)$$

We assign the user n to the corresponding cluster whose center is the closest cluster center to the user n. That is, the user n is equal to $j_n$ cluster, which is

$$j_n = \arg\min_k d_{nk}^c, \quad \forall n \in \mathcal{N}. \quad (15)$$

In addition, it is necessary to select the maximum number of K cluster centers from 2K cluster centers for servicing by the first UAV and the maximum remaining cluster centers to be serviced by the second UAV. The differences between the distances from the center of the k cluster to the first UAV and the second UAV for all ks can be calculated as $\nabla d_k^{cu} = \|w_{u1} - c_k\| - \|w_{u2} - c_k\|$, $\forall k = 1,2,...,2K$.

To maximize the total rate of users served by UAVs, we can sort the distance difference, and a maximum of K from the first least distance difference, - due to the fact that shorter distances mean less interference – is chosen to be served by the first UAV. Also the same as before to satisfy the constraint (8.f), the following inequality must be satisfied.

$$p_u \geq V(d_{in_i^*})^{m+3} \quad (16)$$

$$d_{in_i^*} \leq \sqrt[m+3]{\frac{p_u}{V}} \quad (17)$$

That V remains constant, as the system parameters are determined. Therefore, to maximize the total rate of users served by UAVs, we select the K number of the first least difference of the distances that meet constraint (17) to serve the first UAV, and the remaining K number in that meet constraint (17) Select the applicable for servicing the second UAV.

We define $\mathcal{N}_{ij}$ as a subset of the users in the cluster j (j = 1,2,..., K) that the UAV i (i = 1,2) serves as the center of the cluster. Then, in cluster j, which is served by UAV i, we just need to solve the following problem.

$$\max_{x_i,Y} \sum_{n \in \mathcal{N}_{ij}} x_{in} \left(a \frac{1}{2}\log_2\left(1 + \frac{e}{2\pi}\left(\frac{\zeta p_u h_{1n}}{\sigma_w}\right)^2\right) + b \sum_{k \in \mathcal{N}_{ij}, k \neq n} y_{kn}\right), i = 1,2. \quad (18.a)$$

$$s.t. \quad x_{in} + \sum_{k \in \mathcal{N}_{ij}} y_{nk} \leq 1, \quad \forall n \in \mathcal{N}_{ij}, \quad (18.b)$$

$$\sum_{n \in \mathcal{N}_{ij}} x_{in} = 1, \quad (18.c)$$

$$y_{nk} \leq x_{in}, \quad \forall n, k \in \mathcal{N}_{ij}, n \neq k, \quad (18.d)$$

$$y_{nk} = 0, \quad if\, d_{nk} \geq d_{D2D}, \quad \forall n, k \in \mathcal{N}_{ij} \quad (18.e)$$

$$x_{in} \in \{0,1\}, \quad \forall n \in \mathcal{N}_{ij}, \quad (18.f)$$

$$y_{nk} \in \{0,1\}, y_{nn} = 0, \quad \forall n, k \in \mathcal{N}_{ij}, \quad (18.g)$$

Therefore, we can optimize the UAV allocation to the user, the D2D and $C_K$ allocation (k = 1,2,..., 2K) by optimizing the following problem in (18) for each cluster. Repeat the above operation until the users of the cluster center no longer change.

On the other hand, the following problem in (18), due to its simple limitations, can be solved by a comprehensive search method with little computational complexity. The following problem can be solved as follows.

First, consider $U_n(n\epsilon\mathcal{N}_{ij})$ as the weighted sum of the user n rate associated with the UAV (cluster center among $\mathcal{N}_{ij}$ set users), and the number of users that can be serviced by D2D communication with the cluster center user. We take. Considering the user n as the center of the j cluster, which is served by UAV i, we will have:

$$U_n = (a\, \log_2(1 + \gamma_{in}) + b \sum_{k \in \mathcal{N}_{ij}, k \neq n} y_{kn}) \quad (19)$$

First, we calculate $U_n(n\epsilon\mathcal{N}_{ij})$ according to Equation (19) for all cluster users, and then we sort these values. The maximum value is the optimal value below the problem (18), in cluster j, whose center is served by UAV i. The user corresponding to the maximum value is also selected as the center of the new cluster. For better understanding, the proposed learning-based clustering algorithm is summarized in Algorithm 1. K Select the remaining number that applies to condition (17) to serve the second UAV.

| Algorithm 1 - Cluster-based user association algorithm |
|---|

1. Select 2K points at random as the centers of the clusters. Initialize $C_k$, k = 1, 2, 2K.
2. repeat
   a. Calculate distances from other users to cluster center users based on (14),
   b. Compare each user to a cluster whose center is the nearest cluster center, based on (15) to that user,
   c. From the 2K center of the cluster, select a maximum of K of them, which applies to condition (17), to service the first UAV. Also select the remaining number K, which is valid in condition (17) - to serve the second UAV,
   d. Update the new cluster center user to UAV allocation and user-to-user allocation by optimizing the following (18) for each new cluster,
3. Repeat until the users of the cluster centers do not change.

Since successive solutions (11) and (13) cannot be guaranteed the global optimization, several iterations are required to increase the target value further. Explicitly, by sub-optimal cell allocation, a possible answer to the updated locations of the UAVs can be obtained from (11). Then, the cell allocation can again use the updated locations of the UAVs. Iteration end when the algorithm fails to exceed the target value and the condition (11.b) is met (the lighting requirement is met for all UAV-related users). Otherwise, the algorithm must be run again. We also note that since we broke the main problem into two sub-problems, the final solution is a sub-optimal solution.

## IV. SIMULATION RESULTS

In this section, we review the performance of the proposed method. Assume that N = 200 users are uniformly distributed in a two-dimensional space of $200 \times 200 \ m^2$. The initial horizontal positions and altitude of the two UAVs are assumed to be $w'_{n1} = w'_{n2} = (0,0)m$ and h = 100 m. Each UAV can serve a maximum of K = 8 users directly. In addition, we consider $p_u$ = 200 KW and $d_{D2D}$ = 10 m. In this work, we use LEDs with a relatively wide half-angle, to illuminate a larger area of the ground at a constant height. All system parameters (unless explicitly stated later) are summarized in Table 1. All random results are averaged over a large number of independent replicates.

TABLE I. SYSTEM PARAMETERS

| Parameters | Values |
| --- | --- |
| The size of the area | $200 \times 200 \ m^2$ |
| Refractive index ($n_r$) | 1.5 |
| Semi angle at half power ($\phi_{\frac{1}{2}}$) | 60° |
| Semi angle FOV receiver | 60° |
| Detector area of photodiode (A) | 10 $m^2$ |
| Lighting threshold ($\eta_{th}$) | 0.4 Lux |
| UAV height ($z_u$) | 100 $m$ |
| The nominal power UAV LED is assigned to each related user ($p_u$) | 200 $KW$ |
| power of noise | 0.01 W |
| $d_{D2D}$ | 10 m |
| Illumination target ($\zeta$) | 1 |

As we can see in Figure 2, we can conclude that to maximize the sum rate of users served by UAVs (a>>b), UAVs serve users with less distance and centroids gathering around UAVs, in this case, our method will attention more to sum rate and objective value is increased along with sum rate. In this case, objective value will reach to 43. However, to maximize the number of total D2D connected users (b>>a), centroids is selected in such a way that maximize coverage of connected users to network and centroids Scatter around. In other words, our method concentrate on increasing coverage. In this case, objective value will reach to 51.11.

Figure 3 shows the brightness level of each of the cluster center users connected to the UAVs, as can be seen as the lighting requirement is met.

Then, consider a case in which a = 2/3 and b = 1/3. We need to maximize the total receiving rate of users using UAVs and the total number of users using D2D connectivity. The UAV-user and user-user allocations for this method when K = 8 and K = 6 are shown in Figure 4. From these figures, we can see that users who can serve more users via D2D are served by the nearest UAV.

Figure 5 show that both sum rate and number of D2D connected users are in same order in magnitude and. We notice that with increasing the number of clusters, sum rate and coverage are enhanced and thus finally objective value enhanced about 26.73%.

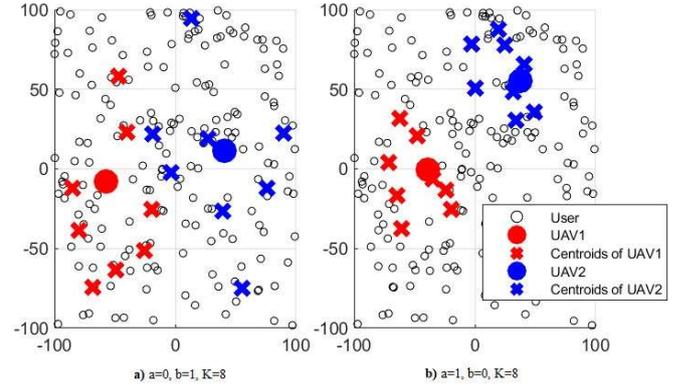

*Figure 2: UAV association to centroids by the proposed method.*

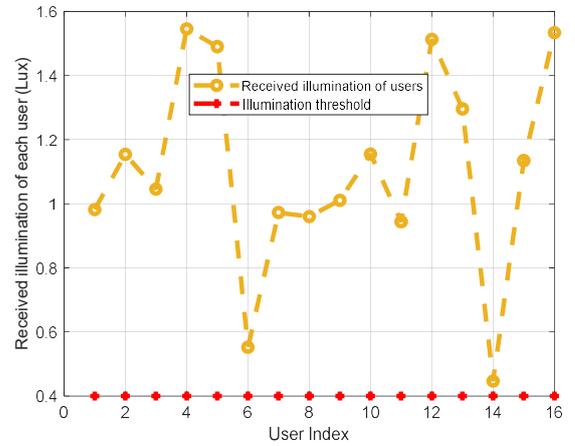

*Figure 3: Brightness level of each user after the last iteration; The system parameters are a = 2/3, b = 1/3, K = 8.*

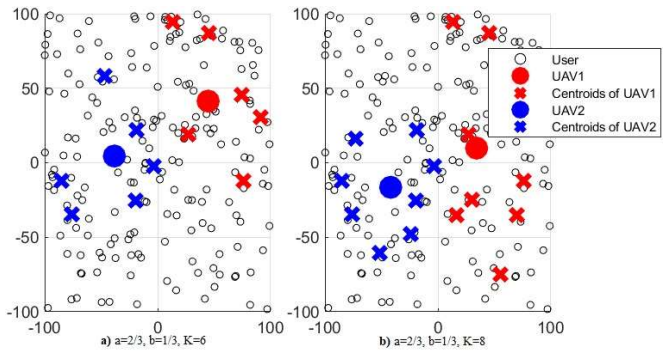

*Figure 4: UAV association to centroids by the proposed method.*

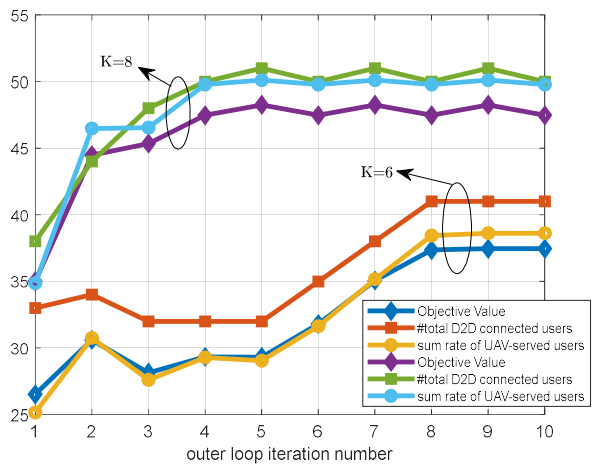

*Figure 5: General convergence of the proposed method, The data rate and received illumination level; with a=2/3, b=1/3*

## V. CONCLUSION

In this paper, we have modeled the two-tiered UAV-assisted VLC network. In this network, each UAV can simultaneously provide communications and illumination for the centroids of ground users over VLC links. Then, in the D2D tier, the centroids retransmit received data from UAV over D2D links to the cluster members. In more details, we have studied the problem of joint user clustering and optimal UAV deployment in the two-tiered UAV-assisted VLC network. To solve this problem, we have presented a two-step method. First, following smallest enclosing disk theorem, we have designed a random incremental construction method to find the optimal UAV locations and meet the minimum brightness requirement of centroids. Then, we have proposed a clustering algorithm inspired by unsupervised learning method to find a suboptimal of user association. Our simulation results show that the proposed scheme on average guarantees the users brightness 0.77 lux more than their threshold requirements. Moreover, the received bitrate under our proposed method is 50.69% more than the scenario in which we have RF Link instead of VLC link and do not optimize UAV location.